\documentstyle[11pt]{article} 
\newtheorem{thm1}{Theorem}[section]

\newcount\exno    
\def\example{\global\advance\exno by 1
\noindent {\bf Example\ \thesubsection.\the\exno.}\quad }
\def\defn{\global\advance\exno by 1
\noindent {\bf Definition\ \thesubsection.\the\exno}}
\def\lemma#1{\global\advance\exno by 1
\noindent {\bf Lemma\ \thesubsection.\the\exno}\quad{\it#1}}

\newcommand{\bg}[1]{\mbox{\boldmath$#1$}}
\newcommand{\bgs}[1]{\mbox{\small\boldmath$#1$}}
\def\esp#1{}
\font\shell=msbm10
\def\tw#1{{\tt #1}}
\def\ep{\epsilon}
\def\bfxi{\bg{\xi}}
\def\bfphi{\bg{\phi}}
\def\D{\mbox{D}}
\def\pr#1{\mbox{pr}^{\mbox{\scriptsize\mit(#1)}\,}}
\def\X{\mbox{\bf v}}
\def\th{^{\mbox{\scriptsize\rm th}}}
\def\pd#1#2{{\partial #1 \over \partial #2}}
\def\biu{\mbox{\boldmath $u$}}
\def\bix{\mbox{\boldmath $x$}}
\def\C{\mbox{\shell C}}
\def\gr{Gr\"obner}
\def\GB{{\sc gb}}

\def\d{\mbox{d}}
\def\i{\mbox{i}}

\def\tx{\textstyle}
\def\tfr#1#2{{\tx{#1\over#2}}}

\def\~#1{\mbox{\protect$\bf\tilde{\mit#1}$}}

\def\Rc{{\cal R}}
\def\N{\mbox{\shell N}}

\def\and{\quad\mbox{and}\quad}

\def\sqr#1#2{{\vcenter{\vbox{\hrule height.#2pt
	      \hbox{\vrule width.#2pt height#1pt \kern#1pt
	      \vrule width.#2pt}
	      \hrule height.#2pt}}}}
\def\square{\mathchoice\sqr34\sqr34\sqr{2.1}3\sqr{1.5}3}

\def\i{\mbox{i}}

\def\PDE{{\sc pde}}
\def\PDES{{\sc pdes}}
\def\pdes{partial differential equations}
\def\ODE{{\sc ode}}
\def\ODES{{\sc odes}}
\def\bigskip{\vspace{\bigskipamount}}
\def\medskip{\vspace{\medskipamount}}
\def\smallskip{\vspace{\smallskipamount}}

\def\cc#1{\kappa_{#1}}
\newcommand{\beq}{\begin{equation}}
\newcommand{\beqn}{\begin{displaymath}}
\newcommand{\bear}{\begin{eqnarray}
\baselineskip=12pt}
\newcommand{\bearn}{\begin{eqnarray*}
\baselineskip=12pt}
\newcommand{\eeq}{\end{equation}}
\newcommand{\eeqn}{\end{displaymath} \noindent}
\newcommand{\eear}{\end{eqnarray}}
\newcommand{\eearn}{\end{eqnarray*}}
\newcommand{\nn}{\nonumber}

\newcommand{\newsection}{\setcounter{equation}{0}
	    \section}
\newcommand{\newsubsection}{\exno=0
	    \subsection}

\begin{document}
\begin{titlepage}
\thispagestyle{empty}
\esp{\begin{frontmatter}}
\title{Algorithms for the Nonclassical Method of Symmetry Reductions}
\author{
\vspace{.5cm}
by \\
\vspace{.5cm}
{\sc Peter A. Clarkson and Elizabeth L. Mansfield} \\
\vspace{.5cm} \\
Department of Mathematics, University of Exeter, Exeter, EX4 4QE,
U.K.\thanks{Email: clarkson@maths.exeter.ac.uk,
liz@maths.exeter.ac.uk}}
\maketitle
\vspace{1in}
\begin{abstract}
In this article we present first an algorithm for calculating the
determining equations associated with so-called ``nonclassical method "
of symmetry reductions
(\`a la Bluman and Cole) for systems of \pdes.
This algorithm requires significantly
less computation time than that standardly used, and avoids
many of the difficulties commonly encountered. The proof of
correctness of the algorithm is a simple application of the
theory of \gr\ bases.

In the second part we demonstrate some algorithms which may be used
to analyse, and often to solve, the resulting systems of
overdetermined nonlinear \PDE s.  We take as our principal example a
generalised Boussinesq equation, which arises in shallow water theory.
Although the equation appears to be non-integrable, we obtain an exact
``two-soliton''
solution from a nonclassical reduction.
\end{abstract}
\end{titlepage}
\parskip6pt
\newfont{\frak}{eufm10}

\newsection{Introduction}
Nonlinear phenomena have many important applications in several aspects
of
physics as well as
other natural and applied sciences. Essentially all the fundamental
equations
of physics are
nonlinear and, in general, such nonlinear equations are often very
difficult to
solve
explicitly. Consequently perturbation, asymptotic and numerical methods
are
often used, with much
success, to obtain {\it approximate\/} solutions of these equations;
however,
there is also much
current interest in obtaining {\it exact\/} analytical solutions of
nonlinear
equations. Symmetry
group techniques provide one method for obtaining such solutions of
\pdes\ (\PDES). These
 have many
mathematical and physical applications, and usually are obtained either
by
seeking a solution in a
special form or, more generally, by exploiting symmetries of the
equation. This
provides a method
for obtaining {\it exact\/} and {\it special\/} solutions of a given
equation in
 terms of solutions
of lower dimensional equations, in particular, ordinary differential
equations
(\ODES). Furthermore
they do not depend upon whether or not the equation is ``integrable''
(in any
sense of the word).

Symmetry groups have several different applications in the context of
nonlinear differential
equations (for further details see, for example, \cite{BK,Olver} and
the references therein):
\begin{itemize}
\item{Derive new solutions from old solutions}. Applying the symmetry
group
of a differential equation to a known solution yields a family of new
solutions
(quite often
interesting solutions can be obtained from trivial ones).
\item{Integration of \ODES}. Symmetry groups of \ODES\ can
be used to reduce the order of the equation (such as to reduce a second
order
equation to
first order).
\item{Reductions of \PDES}. Symmetry groups of \PDES\  are used to
reduce the
total number of dependent and independent variables (for example,
reduce a \PDE\ with two
independent and one dependent variable to an \ODE).
\item{Classification of equations}. Symmetry groups can be used to
classify differential equations into equivalence classes.
\item{Asymptotics of solutions of \PDES}. Since solutions of
\PDES\ asymptotically tend to solutions
of lower-dimensional equations obtained by symmetry reduction, some of
these special solutions will
illustrate important physical phenomena. Furthermore exact solutions
arising from symmetry methods
can often be effectively used to study properties such as asymptotics
and ``blow-up''.
\item{Numerical methods and testing computer coding}. Symmetry groups
and exact
solutions of
physically relevant \PDES\ are used in the design, testing and
evaluation of
numerical algorithms; these
solutions provide an important {practical} check on the accuracy and
reliability of such
integrators.
\end{itemize}

The classical method for finding symmetry reductions of \PDES\ is the
Lie group
method of infinitesimal transformations and the associated determining
equations
 are an
overdetermined, {\it linear\/} system (cf., \cite{BK,Olver}).
Though this method is entirely algorithmic, it often involves a large
amount of tedious algebra and
auxiliary calculations  which can become virtually unmanageable if
attempted manually, and so
symbolic manipulation programs have  been developed to facilitate the
calculations. An excellent survey of the
different packages presently available and a discussion of their
strengths and applications is given
by Hereman  \cite{Here}.

There have been several generalizations of the classical Lie group
method for
symmetry reductions. Ovsiannikov \cite{Ovs} developed the meth\-od of
partially
invariant solutions. Bluman and Cole \cite{BCa}, in their study of
symmetry reductions of
the linear heat equation, proposed the so-called ``nonclassical method
of group-invariant
solutions" (in the sequel referred to as the {\it nonclassical
method\/}), which is also known as
the  ``method of conditional symmetries'' \cite{LWint} and the ``method
of partial
symmetries of the first type'' \cite{Vor}. This method involves
considerably
more algebra and associated calculations than the classical Lie method.
In fact, it has been
suggested that for some \PDES, the calculation of these nonclassical
reductions might be too
difficult  to do explicitly, especially if attempted manually since the
associated determining
equations are now an overdetermined, {\it nonlinear\/} system.
Furthermore, it is known that for
some equations such as the Korteweg-de Vries equation, the nonclassical
method does not yield any
additional symmetry reductions to those obtained using the classical
Lie method, though there are
\PDES\ which possess
symmetry reductions that are {\it not\/} obtained using the classical
Lie group
method. Olver and
Rosenau \cite{ORa,ORb} proposed an extension of the nonclassical method
and
concluded that ``the
unifying theme behind finding special solutions of \PDES\ is not, as is
commonly
 supposed, group
theory, but rather the more analytic subject of overdetermined systems
of
\PDES''.

Clarkson and Kruskal \cite{CK} developed an algorithmic and direct
method for
finding
symmetry reductions (hereafter referred to as the {\it direct
method\/}) and
using it obtained
previously unknown symmetry reductions of the Boussinesq equation. The
novel
characteristic about
the direct method in comparison to those mentioned above, is that it
involves
 no use of
group theory; additionally, for many equations the method appears to be
simpler
to implement than
either the classical or nonclassical methods.

Olver \cite{Olverb} (see also \cite{ABH,Pucci}) has recently shown that
the direct
method is equivalent to the nonclassical
method when the infinitesimals for the independent variables are
autonomous with respect to the dependent variables,
generating a group of ``{\it fibre-preserving transformations}''.

It has been known for several years that there do exist \PDES\ which
possess symmetry reductions
that are not obtained using the classical Lie group method (cf.,
\cite{ORa,ORb}).
Recently the direct and nonclassical methods have been used to generate
many new symmetry
reductions and exact solutions for several physically significant
\PDES,
which represents important progress (cf., \cite{PACrev,Fush} and the
references
therein).

In \S2\ we present an algorithm for calculating the determining
equations for
the nonclassical method for
a system of \PDE.  These are usually  calculated by reducing the
so-called infinitesimal equations, obtained from the group prolongation
of $\Sigma$, with respect
to both $\Sigma$ and the invariant surface conditions $\Psi$.  However,
in
practise, reducing an equation with respect to a system is not
well-defined.  Indeed, unless one
is careful about the choice of term in each equation from which to
back-substitute,
infinite loops can occur in the reduction process.  The theory
used to overcome these difficulties is that of Gr\"obner bases, a
powerful
computational tool in algebra, geometry and logic.
Our essential idea is to first reduce the given system $\Sigma$ using
the invariant surface
conditions $\Psi$ to generate a simplified system $\Sigma_\Psi$. Then
we apply the classical  Lie
method to $\Sigma_\Psi$. We use the theory of Gr\"obner bases, as they
apply to algebraic systems, to provide a proof of correctness of an
algorithm for
finding the determining equations for the nonclassical method
which eliminates the problems and
moreover proceeds efficiently.

In \S2.1 we describe the process of applying the nonclassical
method, and the difficulties encountered, in more detail.  Then we give
an elementary introduction to Gr\"obner bases adapted to our purposes,
and
show why they solve the problems.  Next we give the algorithm for
finding
the determining equations and prove it is correct, followed by some
worked examples that are prototypical.

Since ``nonclassical symmetries" of $\Sigma$
are actually classical symmetries of the system
consisting of $\Sigma$ augmented
by the invariant surface conditions, the basic idea of
using Gr\"obner bases to make the reduction of the
infinitesimal equations well-defined applies to finding
classical symmetries of any system.  We stress that the
``nonclassical symmetries" obtained by the nonclassical method
are {\it not\/} symmetries of $\Sigma$ itself, since they do not
necessarily transform all solutions of the system to other solutions.
Rather, they are symmetries of the augmented system given by
$\Sigma$ together with specified auxiliary conditions.

In \S3 we discuss some algorithms and strategies that have proved
useful in making solving the
determining equations for the nonclassical method, which are
overdetermined and nonlinear,
tractable. We illustrate this with some examples.

The difficulties of the nonclassical method due to the determining
equations being an overdetermined
nonlinear system makes the use of symbolic manipulation  programs more
important. Levi and
Winternitz \cite{LWint} and  Clarkson and Winternitz \cite{CW} in their
applications of the
nonclassical method to the Boussinesq and Kadomtsev-Petviashvili
equations respectively,
interactively  used the {\sc macsyma} program {\sc symmgrp.max}
\cite{CHW}. Nucci \cite{Nucci} has
also developed an interactive program {\sc nusy} in {\sc reduce} for
the
nonclassical method. Here, we use the {\sc maple} package {\tt
diffgrob2}
\cite{MD},
which appears to be the only differential algebra package available
that
can handle equations not solvable for their leading derivative term.
In the appendix we give details of how our algorithm  to obtain
the determining equations
may be implemented using {\sc symmgrp.max}.
The interesting thing here is that we are
using the {\sc symmgrp.max} program for a purpose for which is was not
originally designed,
 however,
it can be adapted to generate the determining equations for the
nonclassical
method since the latter can be interpreted as the determining equations
for
the classical method applied to
an appropriate system of equations.

\newsection{Determining Equations for Symmetries}
\newsubsection{The Classical and Nonclassical Methods}
Suppose one is given a system of \pdes\ \beq
\Sigma=\{f_1=0,\dots,f_r=0\},\label{gpde}
\eeq
where each $f_i$ is some polynomial expression in the
independent variables $\{x_1,\dots,x_n\}$, the dependent
variables $\{u_1,\dots,u_m\}$ and the derivative terms
$\{u_{k,\bgs{\alpha}}\,|\,\bg{\alpha}\in\N^n\}$ where
\beq
u_{k,\bgs{\alpha}}={\partial^{|\bgs{\alpha}|}u_k\over{\partial
x_1^{\alpha_1}
\dots \partial x_n^{\alpha_n}}}.
\eeq
One also can have transcendental and arbitrary functions of the
dependent variables in what follows without affecting the theory.

We recall briefly the method of finding the determining equations for
classical Lie point symmetries thereby fixing our notation.
Let $\biu^{(N)}$ denote the list $u_{k,\bgs{\alpha}}$, where
$k=1,\dots, m$ and
$|\bg{\alpha}|=N$.
The index $\bg{\alpha}+i$ is given by $(\alpha_1,\dots,
\alpha_i+1,\dots, \alpha_n)$,
while $\bg{\alpha}+\bg{\gamma}=(\alpha_1+\gamma_1,\dots,
\alpha_n+\gamma_n)$.

To find the classical Lie point symmetries of the system $\Sigma$, one
takes a group action defined infinitesimally by
\bearn
x^*_i&=x_i+\epsilon \xi_i(\bix,\biu)+O(\epsilon^2),\qquad
i=1,2,\ldots,n,\\
u_j^*&=u_j+\epsilon \phi_j(\bix,\biu)+O(\epsilon^2),\qquad
j=1,2,\ldots,m,
\eearn
where $\bix=x_1,\dots, x_n$ and $\biu=u_1,\dots, u_m$. Then one
requires that this transformation
leaves the set of solutions
$${\cal S}_\Sigma =
\left\{\biu(\bix)|f_1=0,f_2=0,\dots,f_r=0\right\},$$
invariant.

The $N\th$ order partial derivatives transform according to (with
$|\bg{\alpha}|=N$)
\beqn
 {\partial^N u^*_{j} \over \partial
 x_{1}^{*,\alpha_1}x_{2}^{*,\alpha_2}\ldots x_{n}^{*,\alpha_n}} =
 u_{j,\bgs{\alpha}} + \ep \phi_{j}^{[\bgs{\alpha}]}
\left({\bix},{\biu},{\biu}^{(1)},\ldots,{\biu}^{(N)}\right)+O(\ep^2),
\eeqn
where the $N\th$ extension, denoted by
$\phi_{j}^{[\bgs{\alpha}]}\left({\bix},
{\biu},{\biu}^{(1)},\ldots,{\biu}^{(N)}\right)$, is given by the
recursive
formula
\beq
 \phi_{j}^{[\bgs{\alpha}+i]}
 \left({\bix},{\biu},{\biu}^{(1)},\ldots,{\biu}^{(N)}\right) =
{\D\phi_{j}^{[\bgs{\alpha}]}\over \D x_{i}}-
\sum_{\ell=1}^n {\D\xi_\ell\over \D x_{i}}\,{u_{j,\bgs{\alpha}+\ell}}.
\label{recform}
\eeq
 and
\beq
{\D\over \D x_i} \equiv {\partial \over \partial x_i} +
\sum_{\lambda=1}^m\sum_{\bgs{\alpha}}
 u_{\lambda,\bgs{\alpha}+i}  {\partial \over \partial
 u_{\lambda,\bgs{\alpha}}}
\label{totderop}
\eeq
is the {\it total derivative operator\/} \cite{Olver}.
We make the obvious definition
\beq
\D_{\bgs{\alpha}}={\D^{\alpha_1}\over
\D x_{1}^{\alpha_1}}{\D^{\alpha_2}\over \D x_{2}^{\alpha_2}}\dots
{\D^{\alpha_n}\over \D x_{n}^{\alpha_n}}.\label{Dalph}
\eeq

Now consider the system
\beqn
\left\{f_i\left({\bix^*},{\biu^*},{\biu^*}^{(1)}({\bix^*}),\ldots
,{\biu^*}^{(N)}({\bix^*})\right)=0\,|\,
i=1,2,\ldots,r\right\},\label{gpdest}
\eeqn
which is (\ref{gpde}) with ${\biu}$ replaced by ${\biu^*}$ and ${\bix}$
by ${\bix^*}$.
It is easily seen that
\bearn
&&f_i\left({\bix^*},{\biu^*},{\biu^*}^{(1)}({\bix^*}),\ldots
,{\biu^*}^{(N)}({\bix^*})\right)\\
&&\qquad= f_i\left({\bix},{\biu},{\biu}^{(1)}({\bix}),\ldots
,{\biu}^{(N)}({\bix})\right) + \ep\pr{N}{\X}\left(f_i\right) +O(\ep^2),
\eearn
where
\beq
\pr{N}{\X} \equiv \sum_{j=1}^n \xi_j\pd{}{x_j} +
\sum_{k=1}^m\phi_m\pd{}{u_k} + \sum_{k=1}^m\ \sum_{|\bgs{\alpha}|\ge 1}
\phi_{k}^{[\bgs{\alpha}]}\pd{}{u_{k,\bgs{\alpha}}}
\label{prolongN}
\eeq
is known as the $N\th$ {\it prolongation\/} (or $N\th$ {\it
extension\/}) of the
infinitesimal operator
\beqn
{\X} \equiv \sum_{j=1}^n \xi_j({\bix},{\biu})\pd{}{x_j} +
\sum_{k=1}^m\phi_k({\bix},{\biu})\pd{}{u_k},\label{infinopN}.
\eeqn

Let $N$ be the order of the system $\Sigma$.  Requiring that
(\ref{gpde}) is
invariant under the transformation, i.e.
\beq
\left. \pr{N}{\X}\left({f_i}\right)\right|_{{\Sigma}=0}=0,\quad
i=1,\dots, r
\label{prolongeq}
\eeq
 yields an overdetermined, linear
system of equations for the infin\-ites\-imals ${\bfxi}({\bix},{\biu})$
and
${\bfphi}({\bix},{\biu})$, obtained by setting the coefficients of the
different monomials in the derivative terms
$\{u_{k,\bgs{\alpha}}\,|\,1\le k\le m,\ |\bg{\alpha}|\ne 0\}$
in the $\left.\pr{N}{\X}\left({f_i}\right)\right|_{{\Sigma}=0}=0$ to
zero.
This means that the $N\th$ prolongation of
$f_i$, $i=1,2,\ldots,r$, is zero whenever ${\biu}$ is a solution of the
original
system (\ref{gpde}).

The important point to note is that we are considering
the invariance under the group action of the system viewed
as an {\em algebraic system\/} in the indeterminants
$\{x_i,u_j,u_{j,\bgs{\alpha}}\}$, in the relevant jet bundle
(\cite{Olver}, Ch. 2).

The nonclassical method of Bluman and Cole \cite{BCa} for finding
symmetry
reductions
of a system of \PDES\ involves appending the invariant surface
conditions to the system
and finding the classical symmetries of the appended system.
The invariant surface conditions are given by
$$\psi_i\equiv \xi_1 u_{i,x_1}+\xi_2 u_{i,x_2}+\dots+\xi_n
u_{i,x_n}-\phi_i=0,\qquad
i=1,2,\ldots,m. \eqno({\Psi})$$
In this method one requires only the subset of ${\cal S}_\Sigma$ given
by
$${\cal S}_{\Sigma\Psi} = \left\{u(x,t)|f_1=0,f_2=0,\dots,f_r=0,
\psi_1=0,\psi_2=0,\ldots,\psi_m=0\right\},$$ is invariant under the
infinitesimal transformation.
The idea is that since the invariant surface conditions map to
themselves under the prolonged
group action, they are not a restriction
on the infinitesimal equations of the system $\Sigma$,
but rather since $u_{1,x_1}$ and $u_{1,x_2}$,
for example, are no longer independent, the
determining equations will be more general than those for
the classical method.  Hence imposing these conditions
leads to the possibility that there are more solutions, not less.

The usual approach to finding the determining equations for the
nonclassical
 method of the system $\Sigma$ consists of calculating the
 infinitesimal
equations $\pr{N}\X(f)$ for $f\in\Sigma$, where $N$ is the
order of the system $\Sigma$, and reducing them with respect to
the augmented system $G=\Sigma \cup \Psi$.  By reduction with respect
to $G$
is meant elimination (or back-substitution) from $\pr{N}\X(f)$
 of derivatives of one pre-determined term from each
of the equations in $G$. One then reads off the coefficients of the
different monomials in the derivatives of the $u_j$; setting these to
zero are the determining equations.

In practise there are several difficulties.

\example Consider the equation
\beq
\Delta_1\equiv u_{xt}-f(u)=0.\label{eq1}
\eeq
The infinitesimal equation $\pr{2}{\X}\left(\Delta_1\right)$
 (with $x_1=x$, $x_2=t$, $\xi=\xi_1$, $\xi_2\equiv 1$ and
 $\phi=\phi_1$), is
\bearn
\phi^{[xt]}-f'(u)\phi&\equiv&
\phi_{xt}+\phi_{xu}u_t+(\phi_{tu}-\xi_{xt})u_x+(\phi_{uu}-\xi_{xu})u_xu_t\\
&&-\xi_{tu}u_x^2 -\xi_{uu}u_tu_x^2 +(\phi_u-\xi_x)u_{xt}-\xi_tu_{xx}
 -2\xi_uu_xu_{xt}-\xi_uu_tu_{xx}-f'(u)\phi.\label{infeq1}
\eearn
When reducing this equation before reading off the determining
equations, does one reduce the derivative term $u_{xt}$ using
the original equation, or using the $x$ derivative of the invariant
surface condition
\beq
\psi\equiv\xi(x,t,u)u_x+u_t-\phi=0?\label{eq2}
\eeq
The difference of the two reductions is proportional to a differential
consequence
of the system $\{$(\ref{eq1}), (\ref{eq2})$\}$, namely,
\beqn
{\D\over \D x}\psi-\Delta_1\equiv
f(u)+\left(\xi_x+\xi_uu_x\right)u_x+\xi u_{xx}-\phi_x-\phi_uu_x=0.
\eeqn
Using this equation, one can eliminate all $u_{xx}$ terms, given
$\xi\not=0$.  But should
one?
This leads to the next question: ``By which differential consequences,
if
any, do we need to reduce the infinitesimal equations in order to
obtain the determining equations for the nonclassical method?"

\example Second, consider the equation,
\beq
\Delta_2\equiv u_{tt}-u_{xx}=0.\label{eq3}
\eeq
With the same notation as the previous example,
$\pr{2}\X\left(\Delta_2\right)$ is:
\bearn
&&\phi^{[xx]}-\phi^{[tt]}\equiv\\
&&\phi_{xx}+(2\phi_{xu}-\xi_{xx})u_x+(\phi_{uu}-2\xi_{xu})u_x^2-\xi_{uu}u_x^3
+(\phi_u-2\xi_x)u_{xx}-3\xi_uu_xu_{xx}\\
&&-\{\phi_{tt}+2\phi_{tu}u_t-\xi_{tt}u_x+\phi_{uu}u_t^2-2\xi_{tu}u_tu_x
-\xi_{uu}u_t^2u_x+\phi_uu_{tt}-2\xi_tu_{tx}-\xi_uu_{tt}u_x-2\xi_uu_{tx}u_t\}.
\eearn
  Instinctively one would eliminate the $t$
derivatives using the invariant surface condition $u_t=\phi-\xi u_x$,
and then eliminate the $u_{xx}$ terms using $u_{xx}=u_{tt}$.
But eliminating $u_{xx}$ introduces a $u_{tt}$ term, and eliminating
this leads to a $u_{xt}$ term, and eliminating that leads to a
$u_{xx}$ term again.  Clearly care must be taken in the reduction
procedure (by which is meant the successive eliminations or
back-substitutions) to
prevent infinite loops occurring.

These difficulties are all part of the problem of finding the
``normal form" of an equation $f$ with
respect to some given system of equations $G$.
By ``normal form" is meant some well-defined reduction
of $f$ such that no further eliminations from $G$ are possible.

For questions concerning ``normal forms" to be answered in a
well-defined
way, the concept of a Gr\"obner basis is required.
The algorithm to calculate a \gr\ basis for a system of polynomials
over a field was developed by
Buchberger \cite{Bb,Ba} and since then has been extended to a wide
variety of
algebraic scenarios;  the concept
has a large number of applications (see \cite{BCK,Cox}
 and references therein).
In the following section, we discuss Gr\"obner bases as they
are used in our particular application.
\newsubsection{Gr\"obner Bases for Differential Polynomials}

Consider the set
\beqn
\mbox{{\frak Z}}=\{x_i,\xi_{i,\bgs{\delta}},
\phi_{j,\bgs{\delta}},u_{j,\bgs{\alpha}}\,
|\, 1\le i\le n, 1\le j\le m, \bg{\delta}\in\N^{n+m},
\bg{\alpha}\in\N^{n},
\,|\bg{\alpha}|, |\bg{\delta}| \le N\},
\eeqn
where $N$ is some finite number.
The system of equations $\Sigma$,
and the infinitesimal equations obtained by the
prolongation of the group action on $\Sigma$, can be considered to
be polynomials in the elements of {\frak Z} with
complex coefficients.
In this section, we discuss Gr\"obner bases for systems of
polynomials.
We give those definitions required for the sections that follow, and
examples,
relevant to our application.  An excellent introduction to Gr\"obner
bases
can be found in \cite{Cox}.

We denote the set of polynomials in the
indeterminates $\{\zeta_1,\dots,\zeta_s\}$ with coefficients in $\C$ by
${\cal R}=\C[\zeta_1,\dots,\zeta_s]$,
and use the multi-index notation for multiplication
defined for $\bg{\beta}=(\beta_1,\dots, \beta_s)\in \N^s$ by
$\zeta^{\bgs{\beta}}=\zeta_1^{\beta_1}\dots \zeta_s^{\beta_s}$.
Take an ordering ${\cal O}$ on the indeterminates, say
 $\zeta_1<\zeta_2<\ldots <\zeta_s$, and define the lexicographic
ordering lex(${\cal O})$ on the set of monomials
 $\{\zeta^{\bgs{\beta}}\,|\,\bg{\beta}\in \N^s\}$
to be
\beq
\zeta^{\bgs{\beta}}>_{\mbox{lex(${\cal
O})$}}\zeta^{\bgs{\delta}}\label{ord1}
\eeq
if for some $0\le j\le s-1$,
\beq
\beta_s=\delta_s,\ \dots,\ \beta_{s-j+1}=\delta_{s-j+1},\ \
\beta_{s-j}>\delta_{s-j}.
\label{ord2}
\eeq
Following Bayer and Stillman \cite{BS}, another monomial ordering
denoted
here by BS($r$) is given by
$\zeta^{\bgs{\beta}}>_{\mbox{BS($r$)}}\zeta^{\bgs{\delta}}$
if
$\beta_r+\beta_{r+1}+\cdots +\beta_s>\delta_r+\delta_{r+1}+\cdots
+\delta_s$,
 else for some $j$,
$\beta_s=\delta_s,\ \dots,\ \beta_{s-j+1}=\delta_{s-j+1},\ \
\beta_{s-j}<\delta_{s-j}$.
There is a wide variety of monomial orderings available
\cite{refsords}. They are
required to have the so-called {\em compatibility\/}, or
multiplicative, property,  that is,
\beq
\zeta^{\bgs{\beta}}>\zeta^{\bgs{\delta}}\Longrightarrow
\zeta^{\bgs{\gamma}}\zeta^{\bgs{\beta}}
>\zeta^{\bgs{\gamma}}\zeta^{\bgs{\delta}}\qquad \mbox{and}\qquad
\zeta^{\bgs{\gamma}}\zeta^{\bgs{\delta}}>\zeta^{\bgs{\delta}}.
\label{compord}
\eeq

Given an ordering on the indeterminates $\{\zeta_i\}$, we have defined
above several
orderings on the monomials $\{\zeta^{\bgs{\delta}}\}$.
We now discuss orderings on the indeterminates {\frak Z}.

We assume that
$u_{j,\bgs{\alpha}}>\phi_{k,\bgs{\delta}}>\xi_{\ell,\bgs{\delta}}>
x_i$ for all $j$, $k$, $\ell$, $i$, $\bg{\alpha}$ and $\bg{\delta}$,
and require that the
ordering chosen on our indeterminates is compatible with the
differential
structure,  that is, we take the ordering on the indeterminates to
be such that
\beq
u_{j,\bgs{\alpha}}>u_{k,\bgs{\beta}} \Longrightarrow
u_{j,\bgs{\alpha}+\bgs{\gamma}}>u_{k,\bgs{\beta}+\bgs{\gamma}}
\qquad\mbox{and}\qquad
u_{j,\bgs{\alpha}+\bgs{\gamma}}>u_{j,\bgs{\alpha}},
\quad\quad |\gamma|\ne 0 \label{dcom}
\eeq
and similarly for the $\bg{\xi}$ and the $\bg{\phi}$.

  In Example 2.1.2 where the reduction process was infinite, an
  incompatible ordering
had been chosen. In that example we were eliminating $t$-derivatives
using
$\xi u_x+u_t-\phi=0$, which implies $u_t>u_x$.
Then by compatibility (\ref{dcom}), we must have that $u_{tt}>u_{xt}$
and
$u_{xt}>u_{xx}$, so that $u_{tt}>u_{xx}$. Thus using $u_{xx}-u_{tt}=0$
to eliminate occurrences of $u_{xx}$ is an incompatible choice.

Finally, we require that derivatives of the $u$
with respect to one particular pre-chosen
variable, $x_k$ (say), are greater in the order than derivatives with
respect to other
independent variables.  This is
needed to apply the {\em elimination ideals\/} property of \gr\ bases
\cite{Buchelim} in the proof
of correctness of our algorithm.
Thus we want a compatible ordering which is decided first with respect
to
some ordering on the dependent variables, say $u_m>u_{m-1}>\ldots>u_1$,
and then
the number of
derivatives with respect to $x_k$, and then any
compatible choice thereafter.  Such an ordering on the indeterminates
we
will denote by the term $k$-order, or ${\cal O}(k$).  For a $k$-order
${\cal O}(k$) choose $r=r(k)$ such that for
$l\ge r$, $\zeta_l$ represents a derivative term $u_{j,\bgs{\alpha}}$
with $\alpha_k\ne 0$, and for $l<r$, $\zeta_l$ represents either a
derivative term
$u_{j,\bgs{\alpha}}$ with $\alpha_k=0$ or one of the remaining
indeterminates.  We then denote
the monomial ordering BS($r$) by
BS(${\cal O}(k$)).

{\example}
For second order systems of \PDE\ in two independent variables,
suppose we want a compatible ordering
on the derivative terms of $u$ such that $u_t>u_x$.  Two possibilities
are $u_{tt}>u_{xt}>u_{xx}> u_{t}>u_{x}>u$ and
$u_{tt}>u_{xt}>u_{t}>u_{xx}>u_{x}>u$.  The first is
an ordering where  total degree of differentiation
is one of the deciding factors in the
ordering, the second ordering is a $t$-order.

To find the coefficient of a monomial $M$ in $p$,
denoted {\bf coef}$(p,M)$,
one looks for all summands in $p$ of the form $r_i M$ where $r_i\in
\C$.  Then
one defines {\bf coef}$(p,M)= \sum r_i$.  If {\bf coef}$(p,M)\ne 0$,
we
say that $M$ occurs in $p$.
The highest monomial term  occurring in
 a polynomial $p$ is denoted {\sc hmt}$(p)$
and its coefficient is denoted {\bf hcoef}$(p)$.

{\example}  Consider the polynomial
\beqn
p=\xi_u\phi_xuu_{tt}^3u_{xx}u_t+(\xi_{xx}-\xi_u)u_{tt}u_{xx}u_x
+x^2(\phi_u-\xi_x)u_{xt}u_{tt}+u\xi_{xt}u_{xx}u_{tt}u_t^3.
\eeqn
The independent variables are $x$, $t$, the dependent variable is $u$,
the level of prolongation is 2, so that
${\cal O}(x)$ is given initially by $u_{xx}>u_{xt}>u_x>u_{tt}>u_t$, and
${\cal O}(t)$ is given initially by $u_{tt}>u_{xt}>u_t>u_{xx}>u_x$.
The orders ${\cal O}(k)$ are not unique.  Then in the ordering
lex(${\cal O}(x)$), {\sc hmt}($p$)=$u_{tt}u_{xx}u_x\xi_{xx}$, in
lex(${\cal O}(t)$) {\sc hmt}($p$)=$u_{tt}^3u_{xx}u_t\xi_u\phi_xu$
while in BS(${\cal O}(t)$), {\sc
hmt}($p$)=$u_{xx}u_{tt}u_t^3u\xi_{xt}$.

\defn. Suppose for some polynomial $q$ that the {\sc hmt}$(q)$ divides
some monomial M
that occurs in the polynomial $p$, so that $\zeta^{\bgs{\delta}}${\sc
hmt}$(q)=M$,
and that {\bf hcoef}$(q)=a\in\C$. Then the {\it reduction} of $p$ at
$M$ with respect to
$q$ is given by
\beq
p \rightarrow_{q} p-\mbox{{\bf
coef}}(f,M)\,\zeta^{\bgs{\delta}}\,q\left/
a\right. .
\eeq

Thus reduction depends on the ordering used.  The use of
a compatible ordering (\ref{compord}, \ref{dcom})
eliminates the infinite loops observed possible in the
Introduction; see \cite{Ba,MF} where it is proved that with respect to
a compatible
ordering, reduction is a noetherian relation, that is, it must
terminate after
a finite number of steps.

The definition of reduction uses no differentiation, since we wish to
remain within the algebraic domain. Suppose $N$ is the highest
degree of differentiation occurring in the given system $\Sigma$.  When
considering the system $\Sigma \cup \Psi$, the $\Psi$ equations need
to be prolonged to order $N$,  and the system to which our theory
will apply will be
\beq
G=\Sigma \cup \{\D_{\bgs{\alpha}}\psi=0\,|\, \psi\in\Psi,
\bg{\alpha}\in \N^n,\,
|\bg{\alpha}|\le N-1\}
\eeq
where $\D_{\bgs{\alpha}}$ is defined in (\ref{Dalph}).  We assume that
all equations
in $G$ are of the same order of differentiation; if not, those of
lesser order need
to be prolonged.

\defn. A {\it normal form\/} of a polynomial $p$ with respect to a
system
of polynomials $G$, denoted normal$(p,G)$, is achieved when no further
reduction of $p$ with respect to any member of $G$ is possible.

\defn. The {\it ideal\/} generated by a finite
system of polynomials $G\subset \C[\zeta_1, \dots, \zeta_s]$,
 is the set
$I(G)=\left\{\sum_{g\in G} f_{g} g\,|\,  f_{g}\in \C[\zeta_1,\dots,
\zeta_s]
\right\}$.

\defn. A {\it\gr\ basis\/} of an ideal $I(G)$ is a finite set
$H\subset \C[\zeta_1, \dots, \zeta_s]$ such that
$I(H)=I(G)$ and where for all $p\in I(G)$, one has normal$(p,H)=0$.
Thus a \gr\ basis depends upon the ordering used.  We denote a
\gr\ basis
for the ideal generated by $G$ in the monomial ordering {\sc order}
to be \GB($G,\mbox{{\sc order}}$).

Sufficient conditions to characterise a \GB\ and an algorithm to
calculate the
\GB\ for an ideal over a field in any compatible ordering were given by
Buchberger
\cite{Bb}, and it and its generalisations
can be found now in textbooks (see for example \cite{Cox}).
This algorithm has since been implemented in various symbolic
algebra programs, for example in {\sc mathematica} \cite{MAT},
{\sc maple} \cite{Mapleman}, {\sc reduce} \cite{melenk}
or the specialist package {\sc macaulay} \cite{BSM}.

Given a \gr\ basis $G$ of a polynomial ideal $I(G)$, the normal form
of any polynomial  with
respect to $G$ is well-defined.  Let $f$ be a polynomial, and let $h_1$
and
$h_2$ be two normal forms of $f$ with respect to $G$.
Then $h_1-h_2\in I(G)$ by the reduction formula.  Since $G$ is
a \gr\ basis for $I(G)$, $h_1-h_2$ reduces to zero with respect to
$G$.
Since neither $h_1$ nor $h_2$ reduces with respect to $G$, neither can
their difference (reduction must take place at some monomial which must
occur
in at least one of $h_1$ or in $h_2$), so that difference must already
be zero.
Thus different reductions of $f$ with respect to a \gr\ basis are
equal.

{\example}  For the standard example
\beq
u_{xx}=\Delta(x,t,u,u_x,u_t,u_{xt},u_{tt}), \label{stanexam}
\eeq
with invariant surface condition
\beq
\xi u_x +u_t = \phi, \label{invsc}
\eeq
one needs to prolong the invariant surface condition to be of the same
order
as the given equation, namely 2, so one has the system $G$:
\bearn
u_{xx}-\Delta(x,t,u,u_x,u_t,u_{xt},u_{tt})&=0,\\
(\xi_x+\xi_u u_x)u_x+\xi u_{xx}+u_{xt}-\phi_x-\phi_u u_x&=0,\\
(\xi_t+\xi_u u_t)u_x+\xi u_{xt}+u_{tt}-\phi_t-\phi_u u_t&=0,\\
\xi u_x +u_t - \phi&=0.
\eearn
Take the ordering given initially by
$u_{tt}>u_{xt}>u_{xx}>u_t>u_x$.  Finding the \gr\ basis of these
equations is equivalent to considering the first three of them as
equations in $u_{xx}$, $u_{xt}$ and $u_{tt}$ and calculating the
``echelon"
form of the system.
In the ordering given initially by
$u_{tt}>u_{xt}>u_{t}>u_{xx}>u_x$, we eliminate all
the $u_t$ terms using $\xi u_x +u_t - \eta=0$ from the other
conditions,
and then calculate the echelon form of the reduced system to obtain
conditions
for $u_{tt}$, $u_{xt}$ and $u_{xx}$.

To prove the correctness of our algorithm to generate
the determining equations for the nonclassical method, we need the
{\it elimination ideals\/} property of \gr\ bases \cite{Cox,Buchelim}:

\begin{thm1}
Let $\C[\zeta_1, \dots, \zeta_r]$ be the
set of all polynomials over $\C$ in the first $r$ indeterminates.
Suppose $G$ is a \gr\ basis of the
ideal $I(G)$ in the ordering lex(${\cal O}$)
where ${\cal O}$ is $\zeta_1 <\zeta_2<\dots <\zeta_s$ (\ref{ord1},
\ref{ord2}).
Then for for all $1\le r\le s$,
$G\cap \C[\zeta_1, \dots, \zeta_r]$
generates the {\it elimination ideal\/}
$I(G)\cap \C[\zeta_1, \dots, \zeta_r]$.
\end{thm1}

In words, this property means that for every $r$, any polynomial in the
first $r$ indeterminates
that can be found from the generators by addition and by multiplication
by
elements of $\C[\zeta_1, \dots, \zeta_s]$, can be ``read off" from
the \gr\ basis.

For the Bayer and Stillman ordering BS($r$), with $r$ pre-determined,
we have that
$G\cap \C[\zeta_1, \dots, \zeta_r]$ generates
$I(G)\cap \C[\zeta_1, \dots, \zeta_r]$, that is, we obtain only the
one elimination ideal \cite{BS,Cox}.  For our application, we only need
one particular
elimination ideal, that given by using BS(${\cal O}(k$)), while
the Bayer and Stillman orderings are more efficient than the
lexicographic orderings \cite{BS}.

\newsubsection{The Algorithm to calculate the determining equations for
the
nonclassical method.}

Let $\Sigma$ be a system of \PDE\ of ``polynomial type", with $n$
independent variables $\{x_1,\dots,x_n\}$, and $m$ dependent variables
$\{u_1,\dots,u_m\}$.  Terms like
$\sin(u_j)$ or $h(u_j)$ where $h$ is undetermined or arbitrary are
simply considered to be additional indeterminates
in the algebraic computations and do not affect
what follows.

There are $n$ cases to consider, namely where the invariant surface
conditions
($\Psi$) have the forms, for $1\le j\le m$,
\bearn
\xi_1 u_{j,x_1}+\xi_2 u_{j,x_2}+\dots+\xi_{n-1}u_{j,x_{n-1}}+
u_{j,x_n}&=\phi_j,\quad &\mbox{Case}\
n\\
\xi_1 u_{j,x_1}+\xi_2 u_{j,x_2}+\dots+
u_{j,x_{n-1}}&=\phi_j,\quad&\mbox{Case}\
n-1\\
&\vdots&\\
 u_{j,x_1}&=\phi_j,\quad &\mbox{Case}\ 1.
\eearn
That is, we have successively for $1\le k\le n$, that $\xi_{k}$ has
been set
equal to 1 and $\xi_{k+1}= \dots =\xi_n=0$.
We consider each case separately.
In the following algorithm, in the $k{\th}$ case, we require
${\cal O}(k)$ to be a $k$-order, and {\sc order} to be one of
lex(${\cal O}(k$))
or BS(${\cal O}(k$)).
\smallskip

\def\Inf{{\cal I}{\it nf\/}}
\def\RInf{{\cal RI}{\it nf\/}}
\def\DetEqns{{\cal D}{\it et}{\cal E}{\it qns\/}}
{{\noindent{\bf Algorithm: Determining equations for the Nonclassical
Method}
\begin{tabbing}
{OUTPUT}: \=\kill
{INPUT}: \> $\Sigma$, a system of \PDE\ with highest order of
differentiation
$N$;\\
\> $k\in \{1,\dots,n\}$; ${\cal O}(k)$, a $k$-order, {\sc order}$\in \{
$lex(${\cal O}(k$)), BS(${\cal O}(k))\}$.\\
{OUTPUT}: \> $\DetEqns$, the determining equations for the nonclassical
method of the system $\Sigma$\\
\> in the $k{\th}$ case\\
\\
for $j$ from 1 to $m$ let\\
\> $\psi_j:=\xi_1
u_{j,x_1}+\dots+\xi_{k-1}u_{j,x_{k-1}}+u_{j,x_k}-\phi_j$\\
$\Psi^*:=\left\{\D_{\bgs{\alpha}}\psi_j\,|\,1\le j\le m,
\bg{\alpha}\in \N^n,\,|\bg{\alpha}|\le N-1
\right\}$\\
${\cal K}:= \{\mbox{normal}(f,\Psi^*)\, |\, f\in\Sigma \}$\\
${\Inf}:=\{\pr{N}{\X}(f)\, |\, f\in {\cal K}\}$\\
${\cal GB}:=$\GB$({\cal K},\mbox{{\sc order}})$\\
${\RInf}:=\{\mbox{normal}(f,{\cal GB})\, |\, f\in \Inf\}$\\
${\DetEqns}:=\{\mbox{{\bf coef}}(f,u_{1,\bgs{\alpha}_1}\dots
u_{m,\bgs{\alpha}_m})=0
\, |\, f\in \RInf, \bg{\alpha}_j\in \N^{n}\setminus\{0\}\}$\\
end
\end{tabbing}}}

In words, we reduce the given system $\Sigma$ with respect to the
invariant
surface conditions, form the infinitesimal equations of the result,
reduce the infinitesimal equations with respect to an algebraic
Gr\"obner basis of the reduced system, and then read off the
coefficients of the result in the usual manner.

By reducing the equations in $\Sigma$ with respect to $\Psi^*$, that
is,
eliminating all derivatives of the $u_j$ with respect to $x_k$, before
finding the infinitesimal equations, the \gr\ basis calculation is
greatly diminished.  Indeed, for systems consisting of a single
equation,
the calculation is eliminated altogether, and in that case the
algorithm
becomes the classical method, but on the reduced equation.  Since the
algorithm to calculate a \gr\ basis has poor complexity \cite{MM}, our
method is more efficient.

Before discussing the correctness of the algorithm, let us demonstate
it
on Example 2.2.8,

{\example}  Given the general second order equation (\ref{stanexam}),
with invariant surface condition (\ref{invsc}),
the algorithm splits into two cases, $\tau\not\equiv0$ and
$\tau\equiv0$.

\noindent{\bf Case I}. $\tau\not\equiv0$. In this case we set $\tau=1$
(without loss of generality),
and eliminate $u_t$, $ u_{xt}$ and $u_{tt}$ in (\ref{stanexam}) using
(\ref{invsc}), i.e.
\bear \kern-5mm u_t &=&\phi-\xi u_x\label{invsc1}\\
\kern-5mm u_{xt} &=&\phi_x + \phi_uu_x-(\xi_x u_x+\xi_u u_x^2+\xi
u_{xx})\label{invsc1x}\\
\kern-5mm u_{tt} &=&\phi_t + \phi_u(\phi-\xi u_x)
-\xi_tu_x -\xi_uu_x(\phi-\xi u_x) - \xi[\phi_x + \phi_uu_x-(\xi_x
u_x+\xi_u u_x^2+\xi u_{xx})].\label{invsc1t}
\eear  Substituting these into (\ref{stanexam}) yields the \ODE\ (with
$t$ a parameter)
\beq
\tilde\Delta(x,t,u,u_x,u_{xx};
\xi,\xi_x,\xi_t,\xi_u,\phi,\phi_x,\phi_t,\phi_u)=0.\label{redexam}
\eeq  Now apply the classical Lie algorithm to this equation. Thus we
apply the second prolongation
$\pr2\!{\X}$ to (\ref{redexam}) and require that the resulting
expressions vanish for
$u\in\tilde{\cal S}=\left\{u:\tilde\Delta=0\right\}={\cal S}_\psi$,
i.e.,
\beqn
\left.\pr2\!{\X}\left(\tilde\Delta\right)\right|_{\tilde\Delta=0}=0,
\eeqn where
$\xi$ and $\phi$ appear both in (\ref{redexam}) and $\pr2\!{\X}$.
Equating coefficients of powers
of $u_x$ to zero then generates the determining equations.

\noindent{\bf Case II}. $\tau\equiv0$ In this case we set $\xi=1$ and
so the invariant surface condition reduces to
$u_x = \phi(x,t,u)$.  Hence we obtain the differential consequences
\beqn u_{xt} =\phi_t + \phi_uu_t,\qquad
u_{xx} =\phi_x + \phi_uu_x = \phi_x + \phi\phi_x.
\eeqn Substituting these into (\ref{stanexam}) yields the \ODE\ (with
$x$ a parameter)
\beq
\hat\Delta(x,t,u,u_x,u_{t};\phi,\phi_x,\phi_t,\phi_u)=0.\label{goode}
\eeq Now apply the classical Lie algorithm to this equation. Thus we
apply the second prolongation
$\pr2\!{\X}$ to (\ref{goode}) and require that the resulting
expressions vanish for
$u\in\hat{\cal S}=\left\{u(x,t):\hat\Delta=0\right\}$, i.e.,
\beqn
\left.\pr2\!{\X}\left(\hat\Delta\right)\right|_{\hat\Delta=0}=0.
\eeqn
Equating coefficients of powers of $u_t$ to zero then generates the
determining equations.

\noindent{\bf Proof of Correctness:}\quad
We apply the theory described in \S2.1.  We have that $\Sigma$ is a
system of \PDE\ in the dependent variables $\{u_1,\dots, u_m\}$ and
the independent variables $\{x_1,\dots, x_n\}$ of order $N$.
In the $k{\th}$ case, we have
\bearn
\Psi\phantom{^*}&=&\{\xi_1 u_{j,x_1}+\xi_2 u_{j,x_2}+\dots+
u_{j,x_{k}}-\phi_j
\,|\, 1\le j\le m\},\\
\Psi^{*} &=&\{\D_{\bgs{\alpha}}\psi\,|\, \psi\in \Psi,
\bg{\alpha}\in\N^n,\, |\bg{\alpha}|\le N-1\},\\
\Rc\phantom{^*}&=&\C[x_k,u_j,\phi_{j,\bgs{\beta}},
\xi_{j,\bgs{\beta}}\,
u_{j,\bgs{\alpha}}
|\,1\le k\le n,\,1\le j\le m,\,\bg{\beta}\in\N^{n+m},\,
\bg{\alpha}\in\N^n,\,|\bg{\alpha}|,|\bg{\beta}|\le N],\\
\mbox{and}\quad G\phantom{^*}&=&\Sigma\cup\Psi^{*}\subset \Rc.
\eearn
By definition, in the notation used in this article, the
determining equations for the nonclassical method of the system
$\Sigma$ are
\beq
\biggl\{\mbox{{\bf
coef}}\left(\mbox{normal($\pr{N}{\X}$}(f),\mbox{\GB}(\Sigma
\cup \Psi^*)), u_{1,\bgs{\alpha}_1}\dots
u_{m,\bgs{\alpha}_m}\right)=0\,|\, f\in\Sigma,
\bg{\alpha}_i\in\N^n\setminus \{0\}\biggr\}
\eeq
The following Lemma is proved in \cite{Olver}
(Proposition 3.33 and Theorem 3.38).

\lemma{The group prolongations of the invariant surface conditions
satisfy
{\rm pr}\,$\X(\psi)\in I(\Psi^*)$ for $\psi\in  I(\Psi^*)$.}
\xdef\lemi{Lemma \thesubsection.\the\exno}

\lemma{In the $k{\th}$ case of the algorithm, where
$\Psi^{*} =\{\D_{\bgs{\alpha}}\psi\,|\, \psi\in \Psi,
\bg{\alpha}\in\N^n,\, |\bg{\alpha}|\le N-1\}$ and
${\cal K}= \{\mbox{normal}(f,\Psi^*)\, |\, f\in\Sigma \}$, then
$\mbox{{\rm pr}}^{\mbox\scriptsize(N)}\,{\X}g$, for $g\in {\cal K}$
have
all derivative terms of the form $u_{j,\bgs{\alpha}}$ satisfying
$\alpha_k=0$.}
\xdef\lemii{Lemma \thesubsection.\the\exno}

\noindent{\bf Proof:}
Consider the term $\phi_j^{[\bgs{\alpha}]}$.  By examining the formulae
for the
prolongation of the
group action, (\ref{recform},\ref{prolongN}), 
there are two ways that a derivative
of $u_j$ can occur in the $\pr{N}{\X}(g)$:
if either a derivative of $\xi_k$ is non-zero, or if
$g$ has a term of the form $u_{j,\bgs{\alpha}}$ with $\alpha_k\ne 0$.
Since $\xi_k$ is constant, and since no equation in $\Sigma$
after reduction by $\Psi^{*}$ contains a derivative of $u_j$
with respect to $x_k$ (for any $j$), the
lemma is proved.\quad$\square$

Let $f\in \Sigma$ and let $f'$ denote normal($f,\Psi^{*}$).
By \lemi\ and the formula for reduction,
noting that we have {\bf hcoef}($\psi)=1$ for all $\psi\in\Psi^{*}$,
$f'= f+\sum_{\psi\in\Psi^{*}} r_{\psi} \psi$ implying
\beq
\pr{N}{\X}(f')=\pr{N}{\X}(f)+\kappa
\label{redinf}
\eeq
where $r_{\psi}\in \Rc$ and $\kappa\in I(\Psi^*)$.
Hence reducing both sides of (\ref{redinf}) with respect to \GB($\Sigma
\cup \Psi^{*})$ yields
the same result.
 However, in reducing $f$ by $\Psi^{*}$, the
choices of the $r_{\psi}$ have removed the need to reduce the left hand
side of (\ref{redinf}) by $\Psi^{*}$ at all.  This is the content of
\lemii.
Hence there is no need to reduce the elements of $\Inf$ by
the invariant surface conditions.

In $\Psi^{*}$ we have a set of equations for the
$u_{j,\bgs{\alpha}+k}$ occurring linearly.
Now a \GB\ obtained from a reduced set of generators is still a \GB\,
 so that \GB(${\cal K},\mbox{{\sc order}})$ is equal to the elimination
 ideal
\beqn
\mbox{\GB}(G,\mbox{{\sc order}})\ \cap\ \C[u_{j,\bgs{\alpha}}\,|\,
|\bg{\alpha}|\le N,\ \alpha_k=0, 1\le j\le m].
\eeqn
Hence it is sufficient to reduce the elements of $\Inf$ with respect to
\GB(${\cal K},\mbox{{\sc order}})$.
$\square$

Since the system is regarded as an algebraic system on the relevant jet
bundle,
we see why differential consequences obtainable only by further
differentiation, the so-called integrability conditions,
are not relevant.
Of course, one can always investigate the result of reducing with
respect to
additional integrability conditions \cite{PS}.
The following remark by Olver and Rosenau \cite{ORb} is relevant here:
``$\dots$ the reason why Bluman and Cole find nontrivial conditions on
their
groups in order to apply their nonclassical method is that they fail to
take
into account the additional restrictions on the derivatives of $u$
coming
from $\dots$ integrability conditions." According to \cite{ORb}, every
group is a ``weak
symmetry group'' if \underbar{all} the integrability conditions are
taken into account.

Another idea is find the integrability conditions
first and then calculate the determining equations for the enlarged
system,  discussed by Schwarz \cite{Schd} for classical symmetries.
\newsubsection{Examples}
In this section we give some prototypical examples of what our method
looks like in action.
The two examples cover those cases mentioned as being problematic in
the
Introduction.
In the Appendix, we give the input files with which the {\sc macsyma}
package {\sc symmgrp.max} \cite{CHW} calculates the determining
equations for the second example.
Provided at some point a
program written to calculate classical symmetries recognises that the
terms representing infinitesimals in the input equations are the same
functions used by it to represent the infinitesimals, the correct
equations
will be obtained.  For this it may be necessary to input the
infinitesimals
in the internal representation used for them by the package.  Since we
are
using the package for a purpose for which is was not originally
designed, it is
important to know precisely how it can be adapted.

In this section we set $x_1=x$, $x_2=t$, $\xi_1=\xi$ and $\xi_2=\tau$.

{\example The  Nonlinear Wave Equation} $u_{xt}-f(u)=0$.

For the nonlinear wave equation in characteristic coordinates
\beq
u_{xt} = f(u),\label{nwavec}
\eeq
we use the differential consequence of the invariant surface condition
with $\tau\equiv 1$
to eliminate $u_{xt}$ in (\ref{nwavec}) and so obtain
\beq
\phi_x + \phi_uu_x-(\xi_x u_x+\xi_u u_x^2+\xi u_{xx})=
f(u).\label{nwavecieq}
\eeq
Now apply the second prolongation $\pr2\!{\X}$ to this equation and
eliminate $u_{xx}$ using
(\ref{nwavecieq}). Actually, one can only eliminate $\xi u_{xx}$;
if one's program eliminates $u_{xx}$ so that the equation being reduced
is multiplied by $\xi$, that is equivalent to putting $\xi$ in the
coefficient ring
and to assuming that $\xi$ is non-zero.   In some cases it may be
necessary
to consider the case $\xi=0$ separately.  In this case,
allowing $\xi=0$ is equivalent to the next case, $\tau=0$.
Finally, equating coefficients of powers of $u_x$ to zero
yields the following four determining equations:
\bearn
&&\xi{{\xi_{uu}}}-{\xi_{u}^2}=0,\\
&&\xi{\xi_u}{\xi_x}-\xi^{2}{\xi_{xu}}-
 {\xi_{t}}{\xi_u}-{\phi_{u}}\xi{\xi_{u}}+\xi{\xi_{tu}}+{\phi_{uu}}\xi^{2}=0,\\
&&f(u)\xi{\xi_x}-\phi{\phi_{x}}{\xi_u}-f(u)\phi{\xi_u}-{\phi_{x}}{\xi_t}-f(u){\xi_t}+
\phi{\phi_{xu}}\xi-f(u){\phi_{u}}\xi+{\phi_{xt}}\xi+ f'(u)\phi\xi=0,\\
&&\phi{\xi_{u}}{\xi_x}+{\xi_t}{\xi_{x}}-\phi\xi{\xi_{xu}}+
{\phi_{x}}\xi{\xi_u}+3f(u)\xi{\xi_u}-\phi{\phi_{u}}{\xi_u}-
\xi{\xi_{xt}}-{\phi_{u}}{\xi_t}-{\phi_{xu}}\xi^{2}+\phi{\phi_{uu}}\xi+
{\phi_{tu}}\xi=0.
\eearn

In the case $\tau\equiv0$ we set $\xi=1$ and so the invariant surface
condition reduces to
$u_x = \phi(x,t,u)$.
Then we use the differential consequence of this
to eliminate $u_{xt}$ in (\ref{nwavec}) and so obtain
\beq f(u) =\phi_t + \phi_uu_t. \label{nlwtau0}\eeq
Now apply the first prolongation $\pr1\!{\X}$ to this equation and then
eliminate
$u_{t}$ using (\ref{nlwtau0}). This yields one determining equation
\beqn
\phi\phi_{uu}\left(\phi_t-f\right) + \phi_{xu}\left(\phi_t-f\right) -
\phi_u\phi_{xt}
-\phi\phi_u\phi_{tu} - f\phi_u^2+\phi\phi_u{\d f\over\d u}=0.
\eeqn

\example A generalised Boussinesq equation.

Here we derive the determining equations for a ``generalised"
Boussinesq equation
\beq
u_{tt}+u_{xx}+\alpha u_xu_{xt}+\beta u_tu_{xx}+u_{xxxx}=0,\label{pbq}
\eeq
where $\alpha$ and $\beta$ are arbitrary nonzero constants. This
equation, together with
several variants, can be derived from the classical shallow water
theory in the so-called
Boussinesq approximation \cite{Whitham}.
Furthermore the Painlev\'e PDE test due to Weiss, Tabor
and Carnevale \cite{WTC}, suggests that the equation is non-integrable
for any non-zero choice of $\alpha$ and $\beta$.

In the case when $\tau\not\equiv0$, we set $\tau=1$. Using the
invariant
surface condition
$\xi u_x + u_t = \phi$,
we
eliminate $u_{t}$,
$u_{xt}$ and $u_{tt}$ in (\ref{pbq}) to yield
\bear
\phi_t &+&
\phi_u(\phi-\xi u_x) -\xi_{t}u_x -\xi_{u}u_x(\phi-\xi
u_x)-\xi\left[\phi_x +
\phi_uu_x-(\xi_{x} u_x+\xi_{u} u_x^2+\xi
u_{xx})\right]\nn\\ &+&u_{xx}+\alpha u_x\left[\phi_x +
\phi_uu_x-(\xi_{x} u_x+\xi_{u} u_x^2+\xi u_{xx})\right]+\beta
u_{xx}\left(\phi-\xi
u_x\right)+u_{xxxx}=0,\label{pbqdeq}
\eear  Apply the classical Lie algorithm to this equation using the
fourth prolongation
$\pr4\!{\X}$ and eliminate $u_{xxxx}$ using (\ref{pbqdeq}), to yield
the following determining
equations,
\bearn
\xi_{u}=0,\\ \phi_{uu}=0,\\ 2 \phi_{xu} - 3 \xi_{xx}=0,\\
 (\beta + \alpha) (\xi \phi_{u} + 2 \xi \xi_{x} + \xi_{t})=0,\\
 - 2 \beta \xi \phi_{xu} - \alpha \xi \phi_{xu} + \alpha \phi_{u}^{2}
+ \alpha \xi_{x} \phi_{u} + \alpha \phi_{tu} + \beta \xi \xi_{xx}
- 2 \alpha \xi_{x}^{2} - \alpha \xi_{xt}=0,\\
 - \alpha \xi \phi_{x} + 6 \phi_{xxu} + \beta \phi \phi_{u} + \beta
 \phi_{t}
      + 2 \beta \xi_{x} \phi - 4 \xi_{xxx} + 4 \xi^{2} \xi_{x} + 2
      \xi_{x}
      + 2 \xi \xi_{t}=0,\\
 \phi_{xxxx} + \beta \phi \phi_{xx} + \phi_{xx} + \alpha \phi_{x}^{2}
      - 4 \xi \xi_{x} \phi_{x} - 2 \xi_{t} \phi_{x}
      + 4 \xi_{x} \phi \phi_{u} + \phi_{tt} + 2 \phi \phi_{tu}
      + 4 \xi_{x} \phi_{t}=0,\\
 \beta \xi \phi_{xx} - 2 \alpha \phi_{u} \phi_{x} - \alpha \xi_{x}
 \phi_{x}
- 4 \phi_{xxxu}
      - 2 \beta \phi \phi_{xu} - \alpha \phi \phi_{xu} - 2 \phi_{xu}
      + 8 \xi \xi_{x} \phi_{u}\phantom{=0,}\\
 + 2 \xi_{t} \phi_{u} - \alpha \phi_{xt}
      + 2 \xi \phi_{tu} + \beta \xi_{xx} \phi + \xi_{xxxx} + \xi_{xx}
      - 4 \xi \xi_{x}^{2} + 2 \xi_{t} \xi_{x} + \xi_{tt}=0.
\label{pbqnc}
\eearn
In the Appendix we give the {\sc macsyma} input files used to calculate
these determining equations
using the package {\sc symmgrp.max}. We discuss the solution of these
determining equations and the
associated symmetry reductions in Example 3.2.2 below.

In the case when $\tau\equiv0$, we set $\xi=1$. Using the invariant
surface condition
$u_x = \phi(x,t,u)$ we eliminate $u_x$, $u_{xx}$, $u_{xt}$ and
$u_{xxxx}$ in (\ref{pbq}) to yield
\bear u_{tt}&+&\alpha \phi(\phi_t + \phi_uu_t)+(1+\beta
u_t)(\phi_x+\phi\phi_u) \nn\\
&+&\phi_{xxx} + \phi_u\phi_{xx} + 3\phi\phi_x\phi_{uu}+
3\phi_x\phi_{xu} +\phi_x\phi_u^2
+\phi^3\phi_{uuu}\nn\\ &+& 3\phi^2\phi_{xuu} +4\phi^2\phi_u\phi_{uu} +
3\phi\phi_{xxu} +
5\phi\phi_u\phi_{xu}+ \phi\phi_{u}^3 =0.\label{pbqm}
\eear Applying the second prolongation $\pr2\X$ to this equation yields
the following  determining
equations;
\bearn
\phi_{uu}&=&0,\\
\beta\phi_{xx} + (\alpha+\beta)\left(\phi_x\phi_u+\phi\phi_u^2\right) +
(\alpha+2\beta)\phi\phi_{xu} + 2\phi_{tu} &=&0,\\
\phi_{xxxx} + \phi_{xx} + \phi_{tt} + 2\phi\phi_{xu} +
4\phi_{xu}\phi_{xx}+6\phi_x\phi_{xxu}
+4\phi\phi_{uxxx} + 4\phi_x\phi_u\phi_{xu}+ 6\phi\phi_u\phi_{xxu}
&&\\ +
4\phi\phi_u^2\phi_{xu}+ 8\phi\phi_{xu}^2
+(\alpha+\beta)\left(\phi_x\phi_t+\phi\phi_t\phi_u\right)
+\alpha\phi\left(\phi_{xt} +
\phi\phi_{tu}\right)&=&0.
\eearn

\xdef\expbq{\thesubsection.\the\exno}
\newsection{Algorithms for solving the systems of determining
equations}
\newsubsection{Introduction}
Finding the determining equations for the nonclassical method is only
half the story.  Since the systems are overdetermined and nonlinear,
it is necessary to have additional algorithms and strategies to aid in
their
analysis and solution.
In this section we demonstrate
the use of the Kolchin-Ritt algorithm \cite{CF,CMa,MF,Reida,Reidb}
in conjunction with the DirectSearch \cite{CMa} and Reid \cite{RW}
strategies
 to solve the systems of determining equations.

In \S2 of this article, algebraic procedures were used.
The process of calculating algebraic \gr\ bases of successive
prolongations of a system (see for example
\cite{CF}) for the purpose of finding
integrability or compatibility conditions is impractical, even if
one has an {\it a priori} bound on the level of prolongation
necessary to find all such conditions.  Instead, we employ a true
differential analogue of Buchberger's algorithm, with
cross-multiplication
replaced by cross-differentiation, and algebraic reduction replaced
by differential reduction, and so on.  For fully nonlinear systems,
that is, containing equations that are
 not linear in their highest derivative terms, it is necessary
to use pseudo-reduction instead of reduction else the differential
algorithm will not terminate.  The resulting algorithm is called
the Kolchin-Ritt algorithm and has been implemented in packages
in {\sc maple} (\cite{MD,RW}).

In the Kolchin-Ritt algorithm, each pair of
equations in the system given or obtained {\it en route\/}
is cross-differentiated so that their
highest derivative terms become equal.  One then cross-multiplies
by the relevant coefficients  so that these terms cancel.  The result
is
then pseudo-reduced by all known conditions,
and if non-zero, is called an integrability or compatibility
condition of the system.  This process continues
until no new conditions are found.
 In pseudo-reduction one may multiply the
equation being reduced by nontrivial terms in order to effect the
reduction.  A simple example will illustrate our meaning.  Suppose one
wishes to reduce the equation $f:u_{xyy}-u_{xx}u_y=0$ by the equation
$g: u_yu_{xy}-u_{x}^2=0$.  In lex(${\cal O}(x)$), we have that the
highest derivative term in $g$ is $u_{xy}$.  Then a (one-step)
pseudo-reduction of $f$ with respect to $g$ is given by
\beqn
f\rightarrow_{g} u_y f -{\partial g \over \partial y}.
\eeqn
In strict reduction, one is allowed to multiply the equation being
reduced by at most
expressions in the independent variables.
The use of pseudo-reduction means that
the build up of differential coefficients needs to be taken into
account when interpreting the results of the calculation.  Precise
statements describing the output and
 the limitations of the Kolchin-Ritt algorithm
can be found in \cite{MF}, with additional algorithms that eliminate
some (but not all) of the problems.
If the output equations of Kolchin-Ritt are  linear in
their highest derivative terms, then one will have obtained either
the integrability conditions or the differential analogue of the
elimination ideals of the system, depending on whether the ordering
used
is total degree or lexicographic.
In other cases, more work may be required to obtain the
maximum amount of information possible.

The Reid strategy is designed to overcome as far as
possible the problem of the build up of differential coefficients
by adjusting the order in which pairs are chosen to be
cross-differentiated.
In this strategy, the system is divided into four groups of increasing
complexity; single
term equations, linear equations, equations linear in their
highest derivative terms, and fully
nonlinear equations. Into the fourth class is also put those equations
with excessively many terms or high degree of differentiation.
 Each class is reduced with respect to those
in the lower classes.
Beginning with the linear class, the compatibility or integrability
conditions
are calculated and the result used to (pseudo)-reduce the equations
in the higher classes.  This can lead to nonlinear equations becoming
linear for example.  If a condition is found that belongs to a lower
class, the strategy is to recalculate the algorithm on the lower
class with the new condition added.
If a condition is found that belongs to a higher class, the
strategy is to put it there and ``keep it for later". In this way
the need to multiply by differential coefficients in the reduction
process is delayed for as long as possible, and kept to an absolute
minimum.  This strategy has been implemented in {\sc maple} in the
StandardForm procedure of the Reid-Wittkopf Differential Algebra
Package \cite{RW}.
We give an example to demonstrate this strategy, Example
\thesection.3.2.
During the course of the calculations, if any equation is of the
form of an expression raised to a power, we throw away the power
to keep the complexity of the calculation down.  In addition, we
throw away factors known to be non-zero, but we do not throw away
factors involving arbitrary parameters.

The DirectSearch algorithm is a strategy that searches
specifically for conditions less than those in a given set in the
ordering in use.  This strategy was used effectively in \cite{CMa}
to solve a classification problem.  The first example demonstrates
the use of the DirectSearch strategy to  make the calculation of the
Kolchin-Ritt algorithm tractable.

\newsubsection{Examples}
We start with a simple example, namely the determining equations for
the nonclassical method for Burgers' Equation.  We then study a more
complicated example, the generalised Boussinesq equation of Example
\expbq.
All the calculations in this section were performed using the
{\sc maple} package {\tt diffgrob2} \cite{MD}.  By a lexicographic
ordering based on $u_1<u_2\dots <u_m$ and $x_1<x_2<\dots <x_n$ is meant
an ordering on the derivative terms such that $u_{j,\bg{\alpha}}<
u_{k,\bg{\beta}}$ if $j<k$ else $\alpha_n<\beta_n$ else $\alpha_{n-1}<
\beta_{n-1}$ and so on.  It is assumed that $u_{j,\bgs{\alpha}}>x_k$
(of course).
Arbitrary constants of integration in this section are denoted by
$\cc{i}$.

\example The determining equations for the nonclassical method
in the $\tau=1$ case for Burgers' equation
\beq
u_t=u_{xx}+2uu_x
\eeq
are
\bearn
\xi_{uu}&=&0,\\
       \phi_{uu} - 2 \xi_{ux} + 2 \xi \xi_{u} + 2 u \xi_{u}&=&0,\\
 2 \phi \xi_{u} - \phi - 2 \phi_{ux} + \xi_{xx} - 2 \xi_{x} \xi -
 \xi_{x} u - \xi_{t}&=&0,\\
       \phi_{xx} - \phi_{t} - 2 \phi \xi_{x} + u \phi_{x}&=&0.
\eearn
Calculating the Kolchin-Ritt algorithm on this system leads to
expression swell,
so we use a little finesse.
Take a lexicographic ordering based on $\phi>\xi$ and $x>t>u$.
  In this ordering, the
determining equations are listed in ascending order.
The DirectSearch on the first three equations consists of the following
procedure. We cross-differentiate the second and third equations
so that the leading terms cancel, and then reduce
with respect to $\xi_{uu}=0$.  We iterate this procedure recursively
using the result of the previous calculation and the second determining
equation until we have eliminated $\phi$.
We then continue the process using the first determining equation
until we get zero.
This procedure terminates yielding
\beqn
\xi_{u} (\xi_{u} + 1) (2 \xi_{u} - 1)=0.
\eeqn
Thus not only is $\xi_u$ a constant, we know that it can take at most
three
values.  Taking each value
of $\xi_u$ in turn yields three tractable calculations.

\noindent{\bf Case 1}. Reducing the determining equations with respect
to $\xi_u=0$ and performing
the Kolchin-Ritt algorithm yields, after simplification,
\bearn
\xi_u&=&0,\\
\xi_{xx}&=&0,\\
2\xi_{xt}\xi+\xi_{tt}+4\xi_{x}^2\xi+4\xi_{x}\xi_{t}&=&0,\\
\phi+u\xi_u+2\xi\xi_{x}+\xi_{t}&=&0.
\eearn
The equations have been listed in ascending order.  The aim of using a
lexicographic ordering is to try to reduce the integration of the
system to that of a series of {\sc ode} by starting with the least
equation and successively integrating up.
 Setting $\xi=G(t)x+H(t)$ into the third equation and equating
coefficients of powers of $x$ to zero yields
\bearn
G''(t)+6G(t)G'(t)+4G^3(t) &=&0,\\
H''(t)+2G'(t)H(t)+4G^2(t)H(t)+4G(t)H'(t)&=&0.
\eearn
The first can be linearized  to $z'''(t)=0$ by setting
$G(t)=z'(t)/2z(t)$ while the second is linear in $H$ given $G$
so that
$$G(t)={2t+\cc1\over 2(t^2+\cc1 t+\cc2)}\qquad
H(t)=\left\{\cc3\left[{\cc1^2-\cc2\over8(2t+\cc1)}+\tfr14t\right]+\cc4
\right\}G(t).$$
Thus we have $\xi$, and given $\xi$ the fourth equation gives $\phi$.

\noindent{\bf Case 2}. Reducing the original determining equations with
respect to $\xi_u+1=0$
and performing the Kolchin-Ritt algorithm yields
$\phi=0$ and $\xi=-u$, in other words the solution is output!

\noindent{\bf Case 3}. Reducing the second of the determining equations
by $2\xi_u-1=0$
we obtain $\phi_{uu} + \xi + u=0$, implying
$2\phi_{uuu}+3=0$.  Thus $\phi$ is cubic in $u$.
Substituting $\phi=-\tfr14u^3+G(x,t)u^2+H(x,t)u+K(x,t)$ into the
remaining equations, eliminating $\xi$ using
$\phi_{uu} + \xi + u=0$, and setting the coefficients of different
powers
of $u$ to zero yields the system $\Sigma$ for $G$, $H$ and $K$,
\begin{eqnarray}
&& \begin{array}{rcl}
 - G_{t} + G_{xx} + 4 G_{x} G + H_{x}&=&0,\\ K_{xx} - K_{t} + 4 K
 G_{x}&=&0,\\
 H_{xx} + K_{x} - H_{t} + 4 H G_{x}&=&0.\\
\end{array}\label{Sigma}
\end{eqnarray}
In a {\it total degree\/} ordering, the Kolchin-Ritt algorithm on this
system yields no new integrability conditions, and
the formal general solution space appears to depend on six arbitrary
functions.  Since
the system is highly symmetric in its derivative terms, is nonlinear
and is not overdetermined, we need to proceed with caution.
Reverting to a lexicographic ordering, with $K>H>G$ and $t>x$,
cross-differentiating the second and third equations and reducing
yields
$$
 4 K G_{xx} + 2 H_{xxt} - H_{tt} + 8 H_{t} G_{x} + 4 H G_{xt} -
 H_{xxxx}
 - 8 H_{xx} G_{x} - 8 H_{x} G_{xx} - 4 H G_{xxx} - 16 H G_{x}^{2}=0.
$$
Thus, unless $G_{xx}=0$, we have solved for $K$.  We first solve this
singular case by calculating the Kolchin-Ritt algorithm on
(\ref{Sigma})
together with  $G_{xx}=0$ to yield,
\begin{eqnarray}
&&\begin{array}{rcl}
G_{xx}&=&0,\\
- G_{ttt} + 4 G_{xtt} G + 12 G_{xt} G_{t} + 12 G_{x} G_{tt} - 48 G_{x}
G_{xt} G
- 48 G_{x}^{2} G_{t} + 64 G_{x}^{3} G&=&0,\\
H_{x}- G_{t} + 4 G_{x} G &=&0,\\
 - H_{tt} + 8 H_{t} G_{x} + 4 H G_{xt} - 16 H G_{x}^{2}+2 G_{xtt}
- 24 G_{x} G_{xt} + 32 G_{x}^{3}&=&0,\\
K_{x} - H_{t} + 4 H G_{x} + G_{xt} - 4 G_{x}^{2}&=&0,\\
- K_{t} + 4 K G_{x} + G_{tt} - 4 G_{xt} G - 8 G_{x} G_{t} + 16
G_{x}^{2} G&=&0.
\end{array}\label{Singsigma}
\end{eqnarray}
Setting $G(x,t)=xa(t)+b(t)$ in the second equation of (\ref{Singsigma})
and reading off coefficients of $x$ yields,
$$
\begin{array}{rcl}
12 a_{t}^{2} - a_{ttt} - 96 a_{t} a^{2} + 16 a_{tt} a + 64 a^{4}&=&0,\\
4 a_{tt} b - b_{ttt} + 12 a_{t} b_{t} + 12 b_{tt} a - 48 b_{t} a^{2}
- 48 a_{t} a b + 64 a^{3} b&=&0.
\end{array}
$$
The first of these can be linearised to $\zeta''''(t)=0$ by setting
$a(t)=-\zeta'(t)/(4\zeta(t))$, and substituting this into the second
equation yields $(b(t)\zeta(t))'''=0$.  Using this and the final four
equations in the output
(\ref{Singsigma}), which are linear
in $H$ and $K$, yields the singular solution.  This
solution generalises that in \cite{Pucci}.

We return to the general case of (\ref{Sigma}), and use the condition
obtained for $K$ above to eliminate $K$.  Using a Direct Search
method we can solve for $H$ in terms of $G$, and then obtain
two very long expressions for $G$ whose cross-derivative is zero.
None of the singular cases that arise lead to solutions other than that
obtained in the $G_{xx}=0$ case.  This is as far as
elementary differential algebra can take one with (\ref{Sigma}),
and further analysis will depend on additional input, such
as any geometrical information that comes from knowledge of what
the  $G$, $H$ and $K$ actually mean. This could lead to a
transformation in which the analysis is simpler and the geometric
content clearer.  One possibility is the connection, for Burgers'
equation,
between the nonclassical method and the so-called
Singular Manifold method of obtaining solutions using a truncated
Painlev\'e expansion discussed in \cite{Ez}.
This uses an ansatz in which $G$ satisfies a Burgers' type
equation, so that the process of finding the nonclassical
reductions becomes a B\"acklund transformation
for the Burgers' equation.
Another ansatz, used in \cite{ABH}, is $G_{t}=0$.  However, the
full solution appears to be unknown at present.  Further,
the meaning of such a large solution space of nonclassical reductions
appears to be unknown.

\example
In this example we apply the classical and nonclassical
methods for symmetry reductions
of a ``generalised Boussinesq equation" (\ref{pbq}).  The classical
symmetry reductions are calculated not only for comparison purposes,
but to show just how effective the Kolchin-Ritt algorithm can be for
linear systems.  We do not consider the case where either $\alpha$ and
$\beta$ is zero.

\noindent{\bf \thesubsection.\the\exno.A\quad Classical Method}.
Applying the classical Lie
point method to this equation yields the system of linear
overdetermined equations given below
(with $\xi_1\equiv \xi$ and
$\xi_2\equiv\tau$).
\beq
\begin{array}{ll}
\tau_{u}=0, &  \phi_{tu} - \xi_{xt}=0,\\
\tau_{x}=0, &\tau_{t} - 2 \xi_{x}=0,\\
\xi_{u}=0,&\phi_{u} - \tau_{t} + 2 \xi_{x}=0,\\
 \phi_{uu}=0,&(2 \beta + \alpha)\phi_{xu} - \beta \xi_{xx}=0,\\
 2 \phi_{xu} - 3 \xi_{xx}=0,&\beta \phi_{xx} + 2 \phi_{tu} -
 \tau_{tt}=0,\\
(\beta + \alpha) \xi_{t}=0,&\phi_{xxxx} + \phi_{xx} + \phi_{tt}=0,\\
 \alpha \phi_{x} - 2 \xi_{t}=0,&6 \phi_{xxu} + \beta \phi_{t} - 4
 \xi_{xxx} +
2 \xi_{x}=0,\\
4 \phi_{xxxu} + 2 \phi_{xu} + \alpha \phi_{xt} - \xi_{xxxx} - \xi_{xx}
- \xi_{tt}=0.
\end{array}\label{classpbq}
\eeq
Before applying the Kolchin-Ritt algorithm to a linear system,
 the set is simplified to be an ``auto-reduced" set.
In the {\tt diffgrob2} package the procedure used is {\tt reduceall},
while in the Reid-Wittkopf package the procedure is called
{\tt Clearderivdep}.  What these procedures do is
reduce every equation with respect
to every other equation until no further reductions are possible.  With
linear systems this results in no loss of information or solutions.
 The simplification
procedure is also applied to the output of Kolchin-Ritt (or
StandardForm in
the Reid-Wittkopf package).
Using a lexicographic ordering based on
$\xi_2>\phi>\xi_1$ and $x>t>u$,
(recall in these equations, $u$ is an independent variable),
the result of simplification, Kolchin-Ritt and further simplification
appears in the left hand column of Table 1.  Clearly, it is necessary
to run the
calculation again having set $\alpha+\beta=0$.  The result of this
second
calculation appears
 in the right hand column of Table 1.  It will be necessary to
 recalculate
the determining equations with the special values of the parameters
from the
beginning if the package used removes factors in the parameters.
\begin{table}
\beqn
\begin{array}{|r|r|}
\hline
\alpha+\beta\ne 0&\alpha+\beta=0\\ \hline
\begin{array}{r}
\alpha \xi_{xx} (\beta + \alpha)=0\\ (\beta + \alpha) \xi_{t}=0\\
\xi_{u}=0\\
\alpha (\beta + \alpha)\phi_{x}=0\\
\alpha (\beta + \alpha)(\beta \phi_{t} + 2 \xi_{x}) =0\\
 \phi_{u}=0\\
 \tau_{x}=0\\
 \tau_{t} - 2 \xi_{x}=0\\
 \tau_{u}=0
\end{array}&
\begin{array}{r}
 \beta \xi_{xx}=0\\
 \beta \xi_{xt}=0\\
 \beta \xi_{tt}=0\\
 \xi_{u}=0\\
 \beta \phi_{x} + 2 \xi_{t}=0\\
  \beta (\beta \phi_{t} + 2 \xi_{x})=0\\
  \phi_{u}=0\\
  \tau_{x}=0\\
  \tau_{t} - 2 \xi_{x}=0\\
   \tau_{u}=0
\end{array}\\
\hline
\end{array}
\eeqn
\caption{The classical determining equations (\protect\ref{classpbq})
in
Standard Form}
\end{table}
The output is considerably simpler to solve than the original set of
equations,
while the solutions sets are the same. The solutions of these
determining
equations are given in Table 2.

\begin{table}
\beqn
\begin{array}{|r|r|r|}
\hline &\alpha+\beta\ne 0,\alpha\ne 0,\beta\ne
0&\alpha+\beta=0\\ \hline
\xi & \cc1 x+\cc2 & \cc1 x+\cc2 t +\cc3 \\
\tau & 2\cc1 t+\cc3 & 2\cc1 t+\cc4  \\
\phi & -2\cc1 t/\beta + \cc4 & -2(\cc1 t+\cc2 x)/\beta + \cc5 \\
\hline
\end{array}
\eeqn
\caption{Classical Infinitesimals}
\end{table}

We shall now describe symmetry reductions that are generated by these
infinitesimals which
are obtained by solving solving the invariant surface condition
$\xi(x,t,u) u_x + \tau(x,t,u) u_t - \phi(x,t,u) =0$.
In the following $\mu_i$ are constants of integration.

\noindent{\bf Case A.1}. $\alpha+\beta\ne 0$.

\noindent(i) If $\cc1\ne0$, then we set $\cc1=1$ and $\cc2=\cc3=0$, and
obtain the symmetry
reduction
\beq  u(x,t) = w(z) - t/\beta + \tfr12\cc4\ln t, \qquad z=x/t^{1/2},
\eeq where $w(z)$ satisfies
\beq w'''' - \tfr12(\alpha+\beta)zw'w'' - \tfr12\alpha(w')^2  +
\left(\tfr14
z^2+\tfr12\beta\cc4\right)w'' +\tfr34 zw'- \tfr12\cc4=0.
\eeq

\noindent(ii) If $\cc1=0$, then we set $\cc3=1$, and obtain the
symmetry reduction
\beq  u(x,t) = w(z) + \cc4 t,\qquad z = x-\cc2t,
\eeq where $w(z)$ satisfies
\beq w'''' = (\alpha+\beta)\cc2 w'w'' - (1+\cc2^2+\beta\cc4)w'.
\eeq
Setting $W=w'$, this equation can be integrated twice to yield
\beq (W')^2 = \tfr23(\alpha+\beta)\cc2 W^3 -
(1+\cc2^2+\beta\cc4)W^2+\mu_1 W+\mu_0,
\eeq
which is solvable in terms of the Weierstrass elliptic function
$\wp(\theta;g_2,g_3)$ (cf.\
\cite{WW}).

\noindent{\bf Case A.2}. $\alpha+\beta=0$.

\noindent(i) If $\cc1\ne0$, then we set $\cc1=1$ and $\cc3=\cc4=0$, and
obtain the symmetry
reduction
\beq  u(x,t) = w(z) + \left[(\cc2^2-1)t-2\cc2 x\right]/\beta +
\tfr12\cc5\ln t, \qquad
z=(x+\cc2t)/t^{1/2},
\eeq where $w(z)$ satisfies
\beq w'''' + \tfr12\beta(w')^2  + \left(\tfr14
z^2+\tfr12\beta\cc5\right)w'' + \tfr34 zw' -
\tfr12\cc5=0.
\eeq

\noindent(ii) If $\cc1=0$, then set $\cc4=1$, and we obtain the
symmetry reduction
\beq  u(x,t) = w(z) - {2\cc2}\left[xt-\tfr13\cc2t^3-\tfr12\cc3
t^2+\cc5t\right]/\beta,
\qquad z = x-\tfr12\cc2 t^2-\cc3t,
\eeq where $w(z)$ satisfies the linear equation
\beq w'''' = \left[2\cc2z+(2\cc2\cc5-\cc3^2-1)\right]w''-\cc2w' -
2\cc2\cc3/\beta.
\eeq

\noindent{\bf \thesubsection.\the\exno.B\quad Nonclassical Method}.
The determining equations
for the nonclassical method are given in  Example 2.4.2. As in the
classical case, we need to do
the cases $\alpha+\beta\not=0$ and $\alpha+\beta=0$ separately. We
assume $\xi\ne 0$.

\noindent{\bf Case B.1:} $\alpha+\beta\ne 0$.
Take a lexicographic ordering based on $\phi>\xi$ and $x>t>u$.  The
first
step of the Reid strategy involves reducing with respect to small
linear
equations, in this case $2\phi_{xu}-3\xi_{xx}=0$, then calculating
 the compatibility conditions of each pair of determining equations.
This leads to several conditions in $\xi$ alone, including
\beq
(2\beta-3\alpha)(2\beta+\alpha)(3\alpha+4\beta)(\xi_{tt}\xi-2\xi_t^2)=0.
\label{xieqn}
\eeq
The precise coefficients in front of this equation depends on the order
in which pairs are chosen to be cross-differentiated; this affects
which equations appear before others and thus which equations are used
to
do the reduction of new compatibility conditions.
Continuing with the Reid strategy on the equations for $\xi$ yields
for generic $\alpha$ and $\beta$ that
$\xi_{xx}=\xi_{tt}\xi-2\xi_t^2=\xi_{xt}\xi-\xi_x\xi_t=0$.
Choosing all the special values of $\alpha$ for which
(\ref{xieqn}) is zero and redoing the calculation
 leads to the same conclusion so that
\beq
\xi={\sigma x+\cc{3}\over \cc1 t+\cc2},\label{xi:eq2}
\eeq
for all values of $\alpha$ and $\beta$,
where $\sigma$ is either $0$ or $1$.

\noindent{\bf Subcase B.1.1:} $\xi_x\ne 0$.  We set $\sigma=1$.

Reducing the determining equations by (\ref{xi:eq2}) yields one small
equation $\phi_u=(\cc1 -2)\left/(\cc1 t+\cc2)\right.$ and
reducing the remaining equations by this still leaves three
lengthy equations for $\phi$.
In this situation, the DirectSearch strategy is beneficial.
Using a DirectSearch strategy with the equation for $\phi_u$ and
each of the lengthy equations leads to two {\it consistency
conditions\/}
$(\cc1 -2)(\beta\cc1-\beta+2\alpha)=0$ and $\alpha(\cc1 -2)(2\cc1\alpha
-4\alpha+\beta(\cc1^2-5\cc1+2))=0$.
Needless to say, inserting each subcase in the parameters leads to
considerable simplification.
Continuing in this manner we obtain the solutions for each case.
There are two cases which yield solutions.
\begin{itemize}
\item[(i)] For all $\alpha$ and $\beta$, with $\cc1=2$ we obtain the
infinitesimals
\beq
\xi={x+\cc{3}\over 2
t+\cc2},\qquad\phi={-2t+\cc3\over\beta(2t+\cc2)};\label{infs:11}
\eeq these correspond to classical symmetries.
\item[(ii)] If $\beta=2\alpha$ and $\cc1=0$, then we obtain the
infinitesimals
\beq
\xi={x+\cc{4}\over \cc2},\qquad
\phi=-\,{2\alpha u+t\over \alpha\cc2}
+2{(x+\cc{})^2\over\cc2^2\alpha}+\cc3,\label{infs:12}
\eeq which do not correspond to classical symmetries.
\end{itemize}

Setting $\cc2=1/\cc{}$ and $\cc4=0$ in (\ref{infs:12}) yields the
nonclassical reduction
\beq u(x,t) = w(z)\exp (-2\cc{}t) +
\left(\cc{}x^2-t\right)/(2\alpha)+\cc3,\qquad
z=x\exp\left(-\cc{}t\right),\eeq where $w(z)$ satisfies
\beq w'''' = 3\cc{}\alpha zw'w'' + 4\cc{}\alpha ww''+3\cc{}\alpha
(w')^2.
\eeq

\noindent{\bf Subcase B.1.2:} $\xi_x=0$.

In this case we set $\xi={1\left/ ( \cc1 t+\cc2) \right.}$.
Reducing the determining equations with the solution
for $\xi$ and calculating the compatibility conditions according to
the Reid strategy yields $\cc1=0$. If $\alpha\ne \beta$ then unless
$\beta=0$ we obtain
only the trivial solution, $\phi=\phi_0$, a constant.

If $\alpha=\beta$, we can reduce our equations to the simple set
\beqn
\xi=\cc{}, \qquad
\phi_t=\cc{}\phi_x, \qquad \phi_{xxxx}+\phi_{xx}\left\{
\cc{}^2+1+\beta\phi\right\}+\beta\phi_x^2=0.
\eeqn Setting $\phi=\Phi(y)$, where $y=x+\cc{}t$ and $\phi=-12(\Phi+
\cc{}^2+1)/\beta$, the second of the two equations can be written as an
\ODE\ and integrated twice to
yield
\beqn {\d^2\Phi\over\d y^2}=6\Phi^2+ \cc1 y+\cc2.
\eeqn If $\cc1\ne0$ then this equation is equivalent the first
Painlev\'e equation PI (cf.,
\cite{Ince}), whilst if $\cc1=0$ then it is solvable in terms of the
Weierstrass elliptic functions.
Hence we obtain the nonclassical reduction
\beq u(x,t) = v(y)+w(z),\qquad y=x+\cc{}t,\quad z=x-\cc{}t,
\eeq where $v(y)$ and $w(z)$ satisfy
\beq v_{yyyy} + (1+\cc{}^2)v_{yy} + 2\cc{}\beta v_{y}v_{yy} = -
\lambda,\label{veq}\eeq and
\beq w_{zzzz} + (1+\cc{}^2)w_{zz} - 2\cc{}\beta w_{z}w_{zz} =
\lambda,\label{weq}\eeq
respectively, where
$\lambda$ is a ``separation'' constant. Integrating
(\ref{veq},\ref{weq}) and setting $V=v_y$
$W=w_z$, yields
\beq V_{yy} + (1+\cc{}^2)V + \cc{}\beta V^2 = - \lambda y +
\mu_1,\label{Veq}\eeq and
\beq W_{zz} + (1+\cc{}^2)W - \cc{}\beta W^2 = \lambda z+
\mu_2,\label{Weq}\eeq
respectively, where $\mu_1$ and $\mu_2$ are arbitrary constants. If
$\lambda\not=0$ then these
equations are equivalent to the first Painlev\'e equation PI, whilst if
$\lambda=0$ then they are
solvable in terms of Weierstrass elliptic functions. In particular, if
$\lambda=\mu_1=\mu_2=0$,
then equations (\ref{Veq},\ref{Weq}) possess the solutions
\beq V(y) =
-\,{3(1+\cc{}^2)\over2\cc{}\beta}\sec^2\left[\tfr12\sqrt{1+\cc{}^2}\,y\right],
\qquad W(z) =
{3(1+\cc{}^2)\over2\cc{}\beta}\sec^2\left[\tfr12\sqrt{1+\cc{}^2}\,z\right].
\eeq Thus we obtain the exact solution of (\ref{pbq}) with
$\alpha=\beta$ given by
\beq u(x,t) = -\,{3\sqrt{1+\cc{}^2}\over\cc{}\beta}
\tan\left[\tfr12\sqrt{1+\cc{}^2}\,(x+\cc{}t+\delta_1)\right] +
{3\sqrt{1+\cc{}^2}\over\cc{}\beta}
\tan\left[\tfr12\sqrt{1+\cc{}^2}\,(x-\cc{}t+\delta_2)\right].\label{pbqsol}
\eeq

Making the transformation $x\to\i x$, $t\to\i t$ and $u\to\i u$
in (\ref{pbq}) with $\alpha=\beta$ yields
\beq u_{tt}+u_{xx}+\alpha (u_xu_{xt}+
u_tu_{xx})-u_{xxxx}=0.\label{mpbq}
\eeq
Thus from (\ref{pbqsol}) we obtain the exact solution of (\ref{mpbq})
\beq u(x,t) = -\,{3\sqrt{1+\cc{}^2}\over\cc{}\beta}
\tanh\left[\tfr12\sqrt{1+\cc{}^2}\,(x+\cc{}t+\delta_1)\right] +
{3\sqrt{1+\cc{}^2}\over\cc{}\beta}
\tanh\left[\tfr12\sqrt{1+\cc{}^2}\,(x-\cc{}t+\delta_2)\right].\label{mpbqsol}
\eeq
A plot of this solution is given in Figure 1 and a plot of its
derivative with respect to $x$ in
Figure 2. Figure 1 shows that the solution looks like the elastic
interaction of a ``kink'' and an
``anti-kink'' solution. Figure 2 looks like the elastic interaction of
two ``soliton'' solutions.
These are of particular interest since such solutions are normally
associated with integrable
equations, whereas they arise here for a nonintegrable equation.
Furthermore, to our knowledge, this
is the first time that a ``two-soliton'' solution has arisen from a
nonclassical reduction. Normally
such solutions are associated with so-called Lie-B\"acklund
transformations (cf., \cite{AndIb}).

\noindent{\bf Case B.2:} $\alpha+\beta=0$. The analysis of the
determining equations in this case is
similar but more complicated.  We give here only the solution that does
not correspond to a
classical symmetry,
\beqn
\xi=\cc1 t+\cc2,\qquad
\phi= {u\over(t+\cc{0})} + {2(\cc{0}\cc1-\cc2)x +\tfr23\cc1^2t^3
+2\cc1\cc2t^2+2\cc2^2t+\cc3\over\alpha(t+\cc{0})}.
\eeqn Setting $\cc{0}$ we obtain the nonclassical reduction
\beq u(x,t) = tw(z) + \left[6\cc2 x + \cc1^2t^3 -
6\cc2^2t\right]/\alpha +\cc3,\qquad
z=x-\tfr12\cc1t^2-\cc2t,
\eeq where $w(z)$ satisfies
\beq w'''' + \alpha\left[(w')^2-ww''\right] + (1+\cc2^2)w''-3\cc1 w' +
2\cc1^2/\alpha=0.\eeq

It is interesting to note that the values of the parameters for
which nonclassical reductions were
found, not corresponding to classical symmetry reductions,
were precisely $\beta=-\alpha$, $\alpha$ and $2\alpha$. We will not
analyse
the $\tau=0$ case of the determining equations here.

\section*{Acknowledgements}
We thank George Bluman, Willy Hereman, Peter Olver, Greg Reid, Tim
Stokes and Pavel Winternitz for
helpful discussions and the referees for their useful suggestions. We
also thank the Program in
Applied Mathematics, University of Colorado at Boulder, for their
hospitality, where this work was
completed. The support of SERC (grant GR/H39420) is gratefully
acknowledged.  PAC is also grateful
for support through a Nuffield Foundation Science Fellowship and NATO
grant CRG 910729.

\section*{Appendix 1}
We give here the {\sc macsyma} input files used to calculate the
determining equations for
nonclassical reductions, in the $\tau\equiv 1$ case, of the equation
(\ref{pbq}).
They show how to adapt a package designed to calculate classical
symmetries.

The batchfile {\bf pbq1.case} containing the {\sc macsyma} commands to
implement the program {\sc symmgrp.max} is

\begin{verbatim}
batchload("symmgrp.max");
writefile("pbq.out");
/* PERTURBED BOUSSINESQ EQUATION - NONCLASSICAL (TAU = 1) */
/* u_{tt} + u_{xx} + a u_x u_{xt} + b u_t u_{xx} + u_{xxxx} =0 */
batch("pbq1.dat");
symmetry(1,0,0);
printeqn(lode);
save("lodepbq1.lsp",lode);
closefile();
\end{verbatim}

This file in turn batches the file {\bf pbq1.dat} which contains the
requisite
data about (\ref{pbq}):

\begin{verbatim}
p:2$
q:1$
m:1$
parameters:[a,b]$
warnings:true$
sublisteqs:[all]$
subst_deriv_of_vi:true$
info_given:true$
highest_derivatives:all$
depends([eta1,eta2,phi1],[x[1],x[2],u[1]]);
eta2:1;
ut:phi1-eta1*u[1,[1,0]];
uxt:diff(phi1,x[1])+diff(phi1,u[1])*u[1,[1,0]] -
diff(eta1,x[1])*u[1,[1,0]]
    - diff(eta1,u[1])*u[1,[1,0]]**2 - eta1*u[1,[2,0]];
utt:diff(phi1,x[2])+diff(phi1,u[1])*ut - diff(eta1,x[2])*u[1,[1,0]]
    - diff(eta1,u[1])*u[1,[1,0]]*ut - eta1*uxt;
e1:utt+u[1,[2,0]]+a*u[1,[1,0]]*uxt+b*ut*u[1,[2,0]]+u[1,[4,0]];
v1:u[1,[4,0]];
\end{verbatim}

The important thing to note is that {\sc symmgrp.max} recognises that
$\tw{eta1}$ and $\tw{phi1}$
represent the infinitesimals $\xi$ and $\phi$ that are to be
determined. We refer the reader to the
paper by Champagne, Hereman and Winternitz \cite{CHW} for an
explanation of the syntax used and the
purpose of the various commands.

\end{document}